# Anticorrelated Photoluminescence and Free Charge Generation Proves Field-Assisted Exciton Dissociation in Low-Offset PM6:Y5 Organic Solar Cells

*Manasi Pranav, Thomas Hultzsch, Artem Musiienko, Bowen Sun, Atul Shukla, Frank Jaiser, Safa Shoaee, Dieter Neher*

*Affiliations:*

M. Pranav, T. Hultzsch, Dr. A. Shukla, Dr. F. Jaiser, Prof. D. Neher
Soft Matter Physics, Institute of Physics and Astronomy
University of Potsdam
Karl-Liebknecht-Str. 24-25, 14476 Potsdam-Golm, Germany
E-mail: neher@uni-potsdam.de

M. Pranav, B. Sun, S. Shoaee
Optoelectronics of Disordered Semiconductors, Institute of Physics and Astronomy
University of Potsdam
Karl-Liebknecht-Str. 24-25, 14476 Potsdam-Golm, Germany

Dr. A. Musiienko
Department Novel Materials and Interfaces for Photovoltaic Solar Cells,
Helmholtz-Zentrum Berlin für Materialien und Energie,
Kekuléstraße 5, 12489 Berlin, Germany

## Abstract

Understanding the origin of inefficient photocurrent generation in organic solar cells with low energy offset remains key to realizing high performance donor-acceptor systems. Here, we probe the origin of field-dependent free charge generation and photoluminescence in non-fullerene acceptor (NFA) based organic solar cells using the polymer PM6 and NFA Y5 – a non-halogenated sibling to Y6, with a smaller energetic offset to PM6. By performing time-delayed collection field (TDCF) measurements on a variety of samples with different electron transport layers and active layer thickness, we show that the fill factor and photocurrent are limited by field-dependent free charge generation in the bulk of the blend. We also introduce a new method of TDCF called m-TDCF to prove the absence of artefacts from non-geminate recombination of photogenerated- and dark charge carriers near the electrodes. We then correlate free charge generation with steady state photoluminescence intensity, and find perfect anticorrelation between these two properties. Through this, we conclude that photocurrent generation in this low offset system is entirely controlled by the field-dependent exciton dissociation into charge-transfer states.

## Introduction:

Free charge generation in organic photovoltaic cells (OPV) critically relies on photoinduced charge transfer between an electron donor (D) and an electron acceptor (A). This is because the binding energy of excitons in organic semiconductors is much larger than thermal energy at room temperature. To reduce energy losses upon charge transfer (CT), a small energy offset of the frontier orbitals is desired. For electron (hole) transfer, this pertains to a low offset between the lowest unoccupied molecular orbitals (LUMO) (highest occupied molecular orbitals (HOMO)). However, if the energy offset is too small, exciton dissociation at the DA heterojunction becomes inefficient. Numerous studies have shown that an offset range of 0.1-0.5 eV is critical to achieve efficient photon-to-free charge conversion. Inefficient exciton dissociation for low-offset systems has been proven by a relatively lower photovoltaic external quantum efficiency ($EQE_{PV}$), inefficient fluorescence quenching, and reduced free charge generation rates in transient absorption studies. [1]–[6]

More recently, several papers provided evidence for electric-field assisted exciton dissociation in low offset systems[7]–[9]. For example, Weu et al. studied the blend of the donor polymer PffBT4T-2OD with the fullerene-based acceptor $PC_{71}BM$.[8] Compared to the neat blend films, complete devices exhibited a lower steady state photoluminescence (ssPL) intensity, a faster signal decay in transient photoluminescence (TrPL) measurements, as well as more efficient charge generation in transient absorption (TAS). It was concluded that the presence of a built-in electric field in the device assists dissociation of excitons, which would otherwise decay radiatively. Interestingly, these devices exhibited a high fill factor (FF) of 71.4 % despite an active layer thickness of 300 nm. With such a high value of the FF, the photogenerated current saturated within 0.3 V below the open circuit voltage. In other words, a low electric field of ~$10^6$ V.m$^{-1}$ was sufficient to split almost all excitons. This implies a very small exciton binding energy in this system. Note that the use of a different substrate in the full device (ITO) versus the neat film (glass) might affect the blend morphology, but also that dark charge carriers injected by the electrodes may have an additional effect on the excitation dynamics in the device. Liu et al. studied blends of the donor polymer PTQ10 with two non-fullerene acceptors, ZITI-C or ZITI-N.[7] These combinations were chosen because of their different HOMO-HOMO offsets (50 meV for PTQ10:ZITI-C compared to -70 meV for PTQ10:ZITI-N), the latter of which exhibited slower hole transfer and a strong bias-dependence of $EQE_{PV}$. These observations were assigned to field-assisted exciton dissociation. However, the $EQE_{PV}$ alone cannot reveal such a phenomenon since it is the product of several processes that are potentially bias-dependent, including CT state separation and charge extraction, the latter of which may be strongly hindered in the low-donor content blend studied. Nakano et al. investigated photocurrent generation for several DA systems, varying the energy difference between the local singlet exciton (LE) and the CT state (expressed as the driving force $E_G^{opt}$-$E_{CT}$).[9] This study revealed a direct correlation between the efficiency of charge generation and $E_G^{opt}$-$E_{CT}$, highlighting the role of exciton dissociation. Importantly, free charge generation but also the ssPL became more bias-dependent with smaller energy offset. It was proposed that the applied bias promotes the dissociation of the LE for systems with a small driving force. However, a bias dependence of ssPL does not necessarily imply that it originates from field-assisted exciton dissociation. For example, consider a situation where free charge carriers are either extracted to the electrodes (causing an external current) or recombine non-geminately through the reformation of the CT state. Since charge extraction is bias-dependent, so is the repopulation of CT states through non-geminate recombination. For low-offset systems, LE reformation from the CT state becomes efficient, whereby the S1 and the CT state are in dynamic equilibrium.[10], [11] As such, any bias dependence of the extraction-recombination equilibrium may cause equal changes of the LE (and CT) ssPL. More recently, Chow and co-workers performed TrPL and TAS on blends of the acceptor ICPDT-4F with three different donor polymers, thereby tuning the HOMO-HOMO-offset from 430 meV to -50 meV.[12]

While neither the TrPL nor the TAS kinetics was bias dependent for a high energy offset, the blends with small (or even negative) offset revealed an acceleration of the PL decay and an increase of the TAS signal assigned to free charge carriers with increasing reverse bias, and this was explained with an improved CT formation rate. However, none of these studies presented a quantitative comparison between the field-dependence of exciton dissociation and free charge generation, and the contribution of exciton reformation to PL emission was not addressed. Moreover, all these studies concerned only a small electric field range and it remains to be answered whether photocurrent losses even remain at high reverse bias.

Here, we address these issues for the low-energetic offset blend of the donor polymer PM6 and the NFA acceptor Y5. This is a choice model system, due to similarity in chemical and optical properties of the Y5 with the Y6 NFA, which enables a fair benchmark for comparison. Despite this, we find that in contrast to PM6:Y6, the Y5-based blend exhibits a pronounced field-dependence of the photocurrent, meaning that it suffers from voltage-dependent recombination losses. By performing time-delayed collection (TDCF) experiments over a wide bias range, we identify field-dependent free charge generation rather than extraction losses as the leading cause of photocurrent loss. Artefacts due to dark-injected charge carriers[13] or fast recombination at interfaces[14] are ruled out by studying devices with different layer thicknesses and by utilizing a new modified TDCF (m-TDCF) technique. We then show that photoluminescence has the exact opposite bias-dependence as the free charge generation efficiency over the entire voltage range, from $V_{OC}$ to –8V reverse bias, proving that photocurrent generation is entirely governed by field-assisted exciton dissociation in the bulk of the material. At the same time, our study rules out a significant contribution from free charge carrier recombination to the bias-dependent ssPL.

# Results and discussion:

## I. Chemical, Optical and Optoelectronic Properties:

**Figure 1a** shows the chemical structure of the donor polymer PM6 and the NFA Y5, together with the energy levels taken from literature.[3] Compared to Y6, Y5 exhibits a smaller offset of the HOMO to PM6 which is due to the absence of electron withdrawing fluorene units on the terminals of the NFA. **Figure 1b** displays the absorption spectra of films of the neat components and of the PM6:Yx blends. The absorption of neat Y5 is slightly blue shifted compared to that of the popular Y6 NFA, although it is chemically similar in structure to its non-halogenated sibling. This suggests minor differences in intramolecular and intermolecular interactions. It has been shown that mainly molecular aggregates dominate the absorption in Y6 and chemically related Y-series NFAs in neat and blend films.[15] The absorbance of the optimized PM6:Y5 blend is not a perfect superposition of the neat NFA film's absorbance spectra, which we explain with slight differences in the molecular aggregation properties.

Our studies were performed primarily on PM6:Y5 devices in a conventional device geometry, using different electron transport layers for different use cases, which we elucidate in later sections. **Figure S1** (Supporting Information) contains the averaged photovoltaic parameters for these device structures obtained from current-voltage (*JV*) characteristics. Optimized solar cells prepared with PM6:Y5 yield an open circuit voltage $V_{OC}$=0.97 V, a short circuit current density $J_{SC}$=15.5 mA/cm$^2$, and fill factor FF=55%. As expected from the smaller HOMO-HOMO offset, PM6:Y5 exhibits a higher $V_{OC}$ than PM6:Y6. The significantly lower $J_{SC}$ is more surprising, given the similar absorption properties of the PM6:Y5 and PM6:Y6 blend (**Figure 1b**). Finally, the relatively lower FF hints at bias-dependent recombination losses, which are either geminate or non-geminate in nature, or a combination of both.[16], [17] Figure **1d** shows the photovoltaic quantum efficiency ($EQE_{PV}$) as a function of applied bias. As expected of a field-dependent system, a higher negative bias (increasing internal electric field) enhances the $EQE_{PV}$, but the normalized $EQE_{PV}$ spectra overlap perfectly, as seen in **Figure 1d**. In other words, the process dictating field-dependent losses is independent of whether the PM6 or Y5 is initially excited.

To understand the reasons for this bias-dependence, we performed time-delayed collection field measurements (TDCF) on PM6:Y5 and PM6:Y6 devices. TDCF is a powerful optoelectronic transient pump-probe method, which has been used previously to probe the generation, extraction, and recombination of free charge carriers in any solar cell device. The experimental details of the traditional TDCF technique are described in previous works.[14], [18], [19] In essence, a full stack solar cell is held at a bias voltage ($V_{pre}$) during optical excitation with a ≈6 ns laser pulse. After the laser, a negative bias voltage ($V_{coll}$) extracts all free charge carriers with a delay time of $t_{del,coll}$. If $t_{del,coll}$ is very short and a low fluence of optical excitation is chosen, then bimolecular non-geminate recombination of photogenerated charge carriers is nearly absent and the bias-dependence of the extracted charge ($Q_{extr}$) measures the field-dependence of free charge generation. **Figure 1c** shows the results of these measurements for $t_{del,coll}$=1 ns and a fluence of 60 nJ.cm$^{-2}$. It is evident that PM6:Y5 suffers from field-dependent free charge generation, wherein the additional internal field provided by the pre-bias voltage assists in generating more free charge carriers. This is in contrast to PM6:Y6, in which case free charge generation has been shown to be field-independent and barrier-less.[20] The difference between the bias-dependent TDCF and *JV* data at positive applied bias is assigned to non-geminate recombination (NGR), in part due to NGR between photogenerated charges carriers in the devices with dark-injected charge carriers from the electrodes.[14] Notably, this difference is larger for the PM6:Y5 case, shown by the shaded area in Figure **1c**, meaning that NGR is more severe. This will be addressed in a follow-up work.

Here, it should be noted that such fast NGR may also affect the result of TDCF measurements. One probable reason is the above-mentioned pseudo-first order NGR of photogenerated and dark-injected charge carriers.[21] Very high density of dark-injected charge carriers in the vicinity of ohmic contacts accelerates NGR rates, which can result in charge carrier losses even for small delays and fluences. This effect will be particularly pronounced for thin active layers and at $V_{pre}$ close to the built-in voltage of the measured device. In fact, fast NGR in thin polymer:PCBM solar cells was experimentally shown to cause an apparent field-dependence of generation, and this was attributed to surface-related processes.[14] This issue will be addressed in the next two sections.

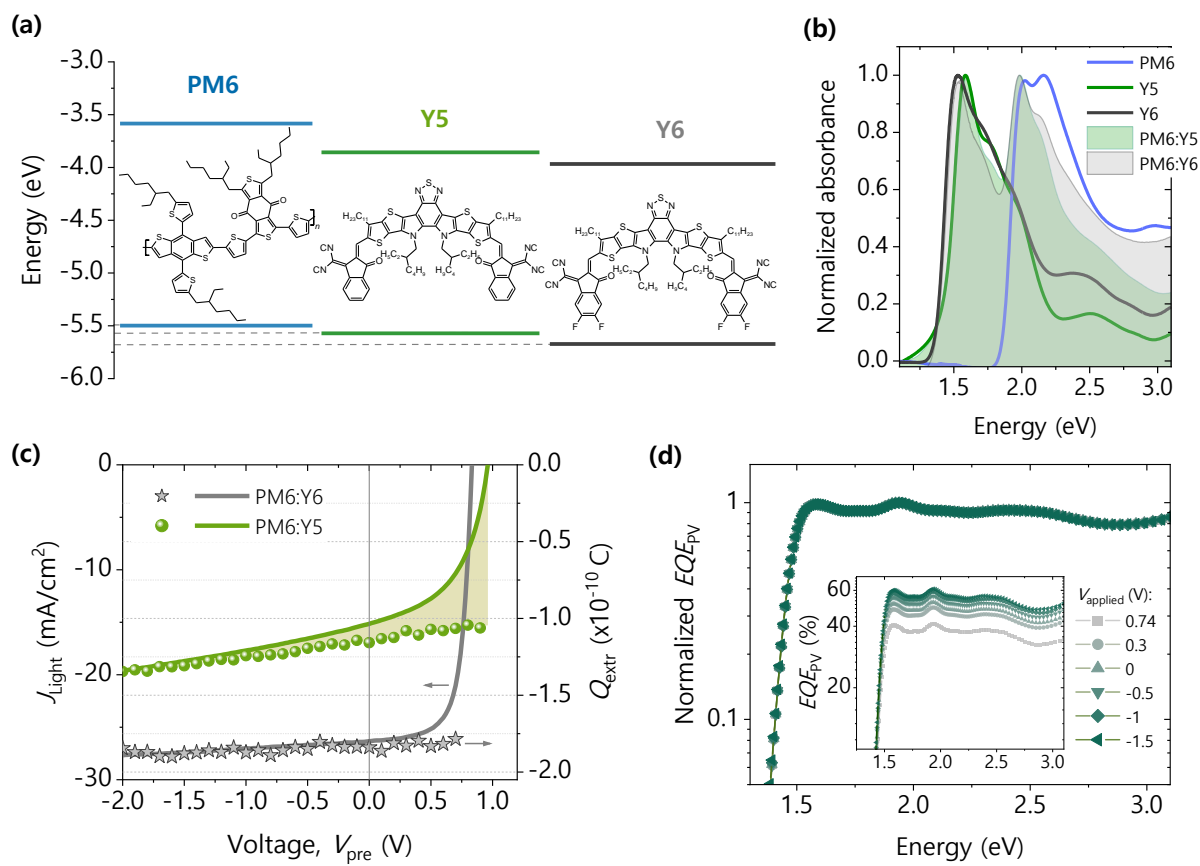

*Figure 1. A view of PM6:Y5 as a model system.* (a) Chemical structures of PM6, Y5 and Y6, and energetics measured by cyclic voltammetry for the three organic molecules.[3] (b) Normalised absorbance of 110 nm neat/blend films from transmission experiments. (c) Overlay of current-voltage (JV) characteristics measured under simulated AM1.5g light (solid line, left axis) and bias-dependent *free charge* formation *from time-delayed collection field (TDCF) (symbols, right axis) for PM6:Y5 and PM6:Y6. In TDCF, the sample is excited with a 60 nJ/cm² fluence laser pulse of 2.33eV and the photogenerated free charge is extracted with a collection bias $V_{coll}$=-2.5 V. (d) Normalised $EQE_{PV}$ as a function of bias for PM6:Y5, showing that the shape of the spectrum is independent of bias. The inset shows the unnormalized spectra.*

## II. Ruling out artefacts in TDCF due to interface-related phenomena: TDCF experiments on different ETLs and sample thicknesses

As a first step to prove that the observed field-dependence of generation originates from processes within the bulk of the active layer, we performed TDCF measurements for (1) a range of active layer thicknesses and by (2) using two different PDI-based ETLs, namely PDINO and PDINN. These different device stacks are illustrated in **Figure 2a**. Given the emphasis placed in these experiments on evaluating the effects of an electric field on generation processes in the active layer, one must be aware of the voltage drops across the different layers of the device. PDINN has been chosen here in comparison to the more common PDINO because its electrical conductivity ($\sigma$) of 6.4x10$^{-3}$ S.cm$^{-1}$ is nearly two orders of magnitude higher than in PDINO ($\sigma$ = 8.1x10$^{-5}$ S.cm$^{-1}$). The electrical conductivities and doping concentrations of both ETLs were obtained using 4-probe Van der Pauw measurements, see Figure **S2** (Supporting information). The higher conductivity of PDINN results in a small voltage drop but also a lower dielectric relaxation time, given by:

$$\tau_d = \frac{\epsilon_0 \epsilon_r}{\sigma} \quad (1)$$

Assuming $\varepsilon_r$=3.5, this yields $\tau_d$ = 48 ps for PDINN versus ca. 6 ns for PDINO, which means faster redistribution of charge carriers within the PDINN layer in response to a change in the applied external bias. This is particularly beneficial for a transient technique such as TDCF where the applied bias is quickly switched from $V_{pre}$ to $V_{coll}$ within a few nanoseconds. On the other hand, a higher free electron concentration of PDINN would lead to a higher dark electron density in the active layer adjacent to the ETL ($N_{d,PDINN}$ = 6x10$^{18}$ cm$^{-3}$ vs $N_{d,PDINO}$ = 2x10$^{17}$ cm$^{-3}$, see supporting information. Further details will be published in a follow-up work). If recombination of photogenerated holes with dark charge carriers affects the TDCF results, this would be visible by the comparison of TDCF data on devices using PDINN *vs* PDINO as the ETL.

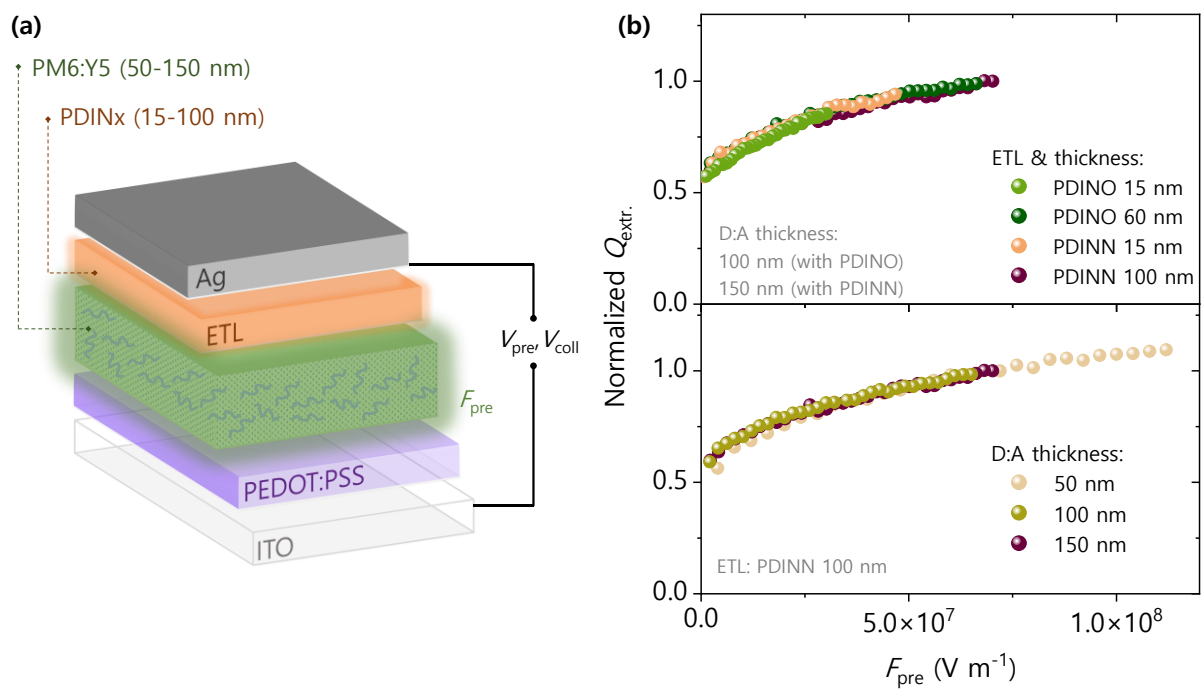

*Figure 2. Impact of the effective field on free charge generation in the bulk of the OPV active layer. (a) Schematic that illustrates conventional PM6:Y5 device structure used in TDCF studies targeted at ruling out the contribution of interfacial phenomena to field-dependent free charge generation. The glow around the ETL and D:A layer signifies the field induced by the pre-bias, $F_{pre}$. By choosing a highly conductive ETL, $F_{pre}$ is maximised across the PM6:Y5 layer which is beneficial for efficient extraction of photogenerated charge carriers. (b) Extracted charge $Q_{extr}$ normalised to a pre-bias field of 7x10$^7$ Vm$^{-1}$ for (top panel) different ETL properties and (bottom panel) various PM6:Y5 layer thicknesses.*

The $Q_{extr}$ against the pre-bias field ($F_{pre}$) for each device structure, normalised to $F_{pre}$=7x10$^7$ Vm$^{-1}$, is plotted in the top and bottom panels of Figure **2b**. $F_{pre}$ is calculated using the built-in voltage $V_{bi}$ of the measured OPVs given by $F_{pre} = \frac{V_{bi} - V_{pre}}{d}$ (See Figure S3 for the $V_{OC}$ and $V_{bi}$ values). Particularly good agreement of the data across all device structures implies that the field-dependence of extracted free charge generation remains unchanged regardless of the active layer thickness and ETL, although PDINO and PDINN possess largely different transport properties. This gives sound evidence that the observed field-dependence of free charge generation is indeed a bulk phenomenon. Also, the overlap of data in the top panel of Figure **2b** means that the TDCF results are not altered by the higher doping concentration of PDINN or the slower response of the PDINO layer when applying the collection bias. We note that the field-dependence of free charge generation is gradual over the entire field range, characterized by a lack of saturation of $Q_{extr}$ even at high field conditions above $1\times10^8$ Vm$^{-1}$. This is even more clearly seen in the semi-log representation of Figure **2b** in Figure **S3 (SI).** Interestingly, this observation deviates from the predictions of the Onsager-Braun model for CT separation, from which we expect a more sudden transition to the saturation regime for devices with fairly efficient zero-field charge separation, as seen here.[22]

## III. Ruling out artefacts in TDCF due to dark-injected charge carriers: Modified TDCF

While the above data provides compelling evidence that field-dependent free charge generation originates in the bulk of the PM6:Y5 layer, we went one step further to specifically address the role of dark-injected charge carriers in the TDCF data. For this, a modified version of the TDCF technique is herein introduced (m-TDCF). Figure **3a** succinctly illustrates different TDCF parameters and temporal characteristics of the photocurrent transients in both traditional- and m-TDCF measurements. In traditional TDCF, the voltage transient returns to $V_{pre}$ after extracting photogenerated charge carriers with $V_{coll}$. This means that for typical extraction times of 2 to 10 µs and a repetition rate of 500 Hz, the OPV is held at $V_{pre}$ for almost 2 ms before being optically excited once again by the next laser pulse. This period is more than sufficient to establish steady-state conditions at the metal-semiconductor interface.

To cope with this, we have altered the shape of the applied voltage transient waveform. Within the same frequency of the laser pulse, the voltage transient in m-TDCF returns to 0 V instead of to $V_{pre}$ after completion of the extraction time, as shown in Figure **3a**. The voltage is then switched to $V_{pre}$ for a short duration prior to and during illumination by the next laser pulse. The idea is to limit the duration of $V_{pre}$ to just enough time for the OPV as a capacitor to become fully charged, but well before a steady state dark condition is established. In Figure **3a**, the delayed application of $V_{coll}$ after the laser pulse is denoted by the parameter $t_{del,coll}$, and the time at which $V_{pre}$ is applied prior to the laser pulse by the parameter $t_{adv,pre}$. To determine the optimum value for $t_{adv,pre}$, we take into account both the RC time of the OPV and the internal latency of the function generator that applies the voltage transient. For the measured PM6:Y5 OPVs, a pre-bias duration of $t_{adv,pre}$=24 ns is well suited for an active layer thickness of 110 nm, for which we found the RC time to be ca. 7 ns. The criteria to determine the optimal value for $t_{adv,pre}$ in this device are described in the Supporting information and in Figure **S4**.

The upper panel of Figure **3b** shows $Q_{extr}$ as a function of $V_{pre}$ for $t_{adv,pre}$=24 ns and different $t_{del,coll}$ using the m-TDCF technique. We find that the bias-dependence of $Q_{extr}$ is the same for $t_{del,coll}$=1 ns and $t_{del,coll}$=10 ns. This rules out fast NGR losses prior to charge carrier extraction. We increased $t_{del,coll}$ even further to demonstrate the efficacy of the modified voltage waveform on reducing NGR near the electrodes. This data, shown for m-TDCF in Figure **3b,** is compared with that from traditional TDCF in SI Figure **S5**. As expected for longer $t_{del,coll}$, there are increasing charge carrier losses due to NGR among photogenerated charge carriers but also between photogenerated and dark-injected charge carriers. Eventually at long enough extraction delays, the $Q_{extr}$ from both TDCF techniques approaches the JV characteristics. Notably, NGR losses at longer $t_{del,coll}$ are enhanced in traditional TDCF compared to m-TDCF even for fairly fast collection ($t_{del,coll}$=10 ns) (see SI Figure **S5**). This proves that the modified waveform in m-TDCF certainly reduces the presence of dark injected charge carriers available for NGR, by shortening the application of $V_{pre}$. We observe, importantly, that under optimum pre-bias delay and immediate extraction after laser excitation, the free charge generation recorded from traditional and m-TDCF overlap perfectly, seen in the upper panel of Figure **3b**. This confirms that traditional TDCF measurements on PM6:Y5 are indeed not affected by the injection of dark charge carriers over the entire voltage range (including at positive $V_{pre}$).

Lastly, we note that if the delay between pre-bias application and photoexcitation is shorter than necessary to ensure the complete build-up of $V_{pre}$ over the active layer, the apparent field-dependence would be reduced. Indeed, by reducing the duration of $t_{adv,pre}$ from 24 ns to 9 ns and collecting the photogenerated charge carriers immediately after the laser pulse, the field-dependence of free charge generation appears diminished, as seen by comparing the purple stars with the ideal delay condition in the lower panel of Figure **3b**.

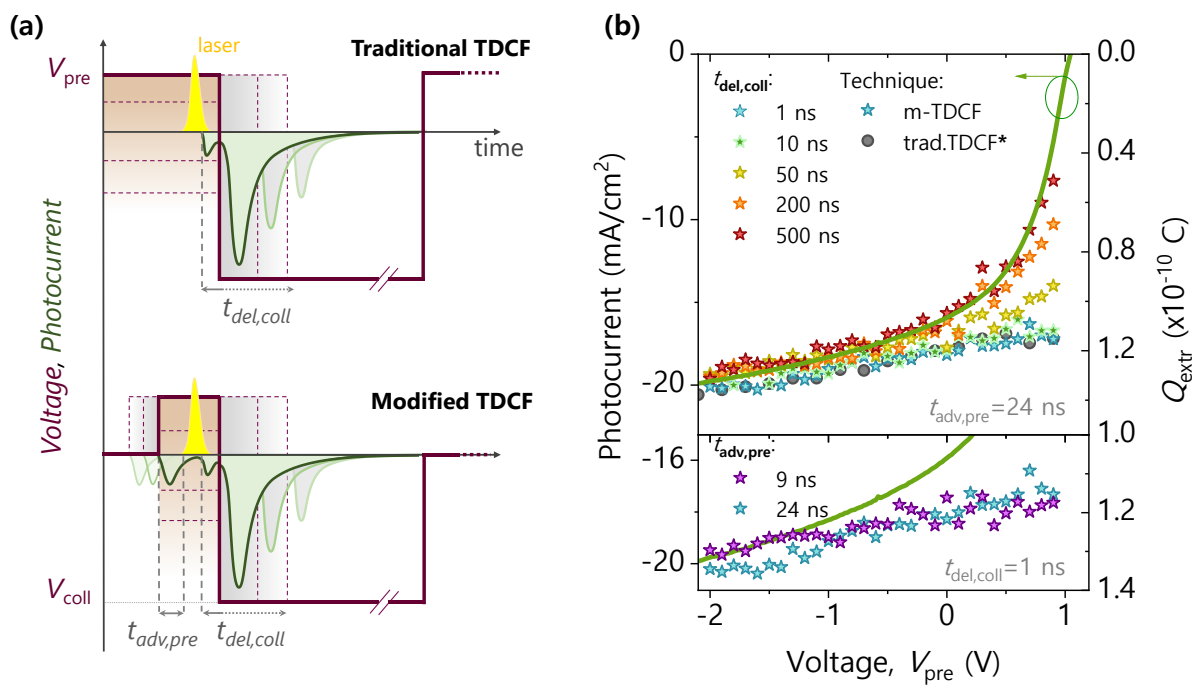

***Figure 3. Modified-TDCF (m-TDCF) measurements on PM6:Y5 for different delay settings.** (a) An illustration of the temporal characteristics of both the traditional and the newly developed m-TDCF techniques. Voltage and photocurrent transients are shown in plum and green, respectively. The dotted lines depict voltage ranges and time scales of applying the pre-bias $V_{pre}$ and collection bias $V_{coll}$. In contrast to traditional TDCF, the voltage waveform in m-TDCF returns to 0 V after complete charge extraction and switches to $V_{pre}$ only shortly before the next laser pulse. Thereby, an additional temporal parameter $t_{adv,pre}$ defines how long the OPV is charged with $V_{pre}$ prior to optical excitation. For different durations of $t_{del,coll}$ and $t_{adv,pre}$, the respective transient photocurrents are illustrated as translucent*

*curves. (b) Measurement of m-TDCF on PM6:Y5 with a PDINN ETL. The photocurrent of PM6:Y5 from JV is overlaid on bias-dependent extracted charge obtained with m-TDCF, for a 2.33 eV laser excitation of 60 nJ/cm$^2$ fluence. The top and bottom panels show $Q_{extr}$ for various $t_{del,coll}$ and $t_{adv,pre}$, respectively. Grey circles depict the bias-dependence of the extracted free charge measured with traditional TDCF (\*with $t_{del,coll}$ =1 ns). A comparison is drawn to m-TDCF, where the collection delay is also 1 ns, and $t_{adv,pre}$=24 ns.*

## IV. Interplay of photoemission and free charge generation:

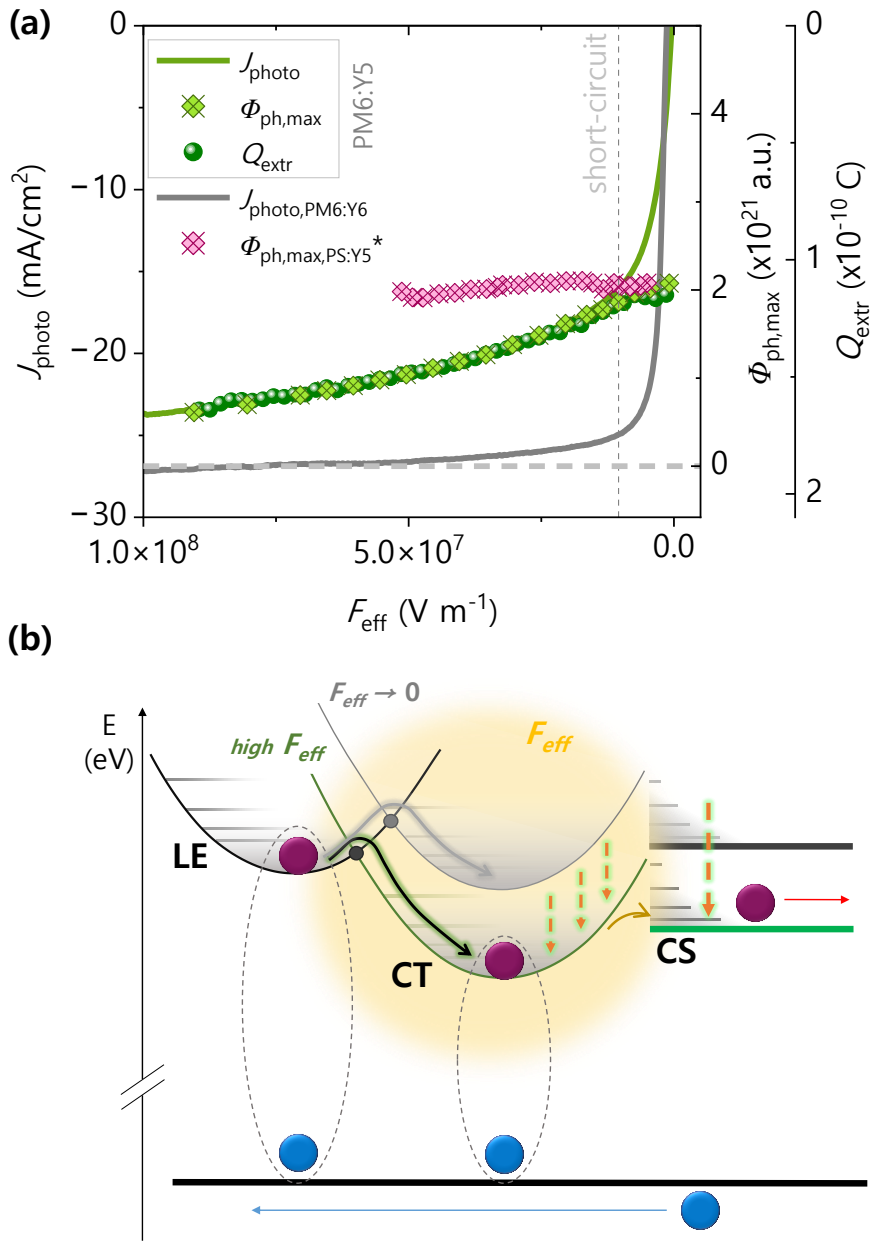

***Figure 4. Correlating photoluminescence and generation in PM6:Y5, benchmarked against PM6:Y6.**
(a) Overlay of the photocurrent ($J_{photo}$), $Q_{extr}$ from traditional TDCF and the peak ssPL photon flux $\Phi_{ph,max}$, all plotted as a function of the internal electric field for a PM6:Y5 devices with a 110 nm active layer and a 15 nm PDINN ETL. Also shown is $J_{photo}$ for PM6:Y6, and $\Phi_{ph,max}$ as function of bias for a PS:Y5 blend* with the same device structure (*the $\Phi_{ph,max}$ for PS:Y5 is scaled such that the emission intensity at $V_{OC}$ matches that of the blend devices). The vertical dotted line marks the effective field $F_{eff}$ at short-circuit conditions. (b) Marcus-type presentation of the potential curves of the local singlet exciton (LE) and the CT state. The application of an electric field (depicted by the yellow halo) reduces the energy of the CT state and thereby diminishes the barrier for the LE to CT transition. For simplicity, the energy of the CS state (and the CS state lowering due to the electric field) is indicated by horizontal lines.

TDCF measurements on PM6:Y5 with different device structures & thicknesses and different voltage transients (TDCF vs m-TDCF), described earlier, provide comprehensive evidence for a pronounced field-dependence of free charge generation in the bulk of the active layer. This raises the question whether it is inefficient (field-dependent) LE dissociation that mainly limits free charge efficiency. To this end, we measured the intensity of steady-state photoluminescence as a function of bias, covering a wide electric field ($F_{eff}$) range up to $9 \times 10^7$ V.m$^{-1}$. To benchmark the emission characteristics of PM6:Y5, we also measured the field-dependent ssPL of Y5 in an inert polystyrene (PS) polymer matrix. The comparison of the ssPL spectra of the PM6:Y5 blend with that of PS:Y5 shows that the blend emission is entirely controlled by the radiative decay of the Y5 LE for the same excitation wavelength (see Figure **S6a** in the SI). Also, the applied bias does not affect the shape of the ssPL spectrum (see Figure **S6b** in the SI). We find foremost that the peak photon flux ($\Phi_{ph,max}$) emitted from the PM6:Y5 device diminishes with increasingly negative bias (increasing electric field), suggesting that a stronger $F_{eff}$ depletes populated LE states as emission channels, shown in Figure **4a**. This is a clear indication of field-assisted exciton dissociation. However, given the smaller HOMO offset in PM6:Y5, LE reformation via CT back transfer may serve as a competing pathway to LE dissociation. To assess the contribution of radiative decay from reformed LE excitons in the ssPL spectra, we recorded the PL quantum efficiency (PLQY) of the acceptor in PS and EL quantum efficiency (ELQY) of the blend on full stack devices. The obtained values of PLQY$_{PS:Y5}$=2.45x10$^{-2}$ and ELQY$_{PM6:Y5}$=2.2x10$^{-3}$ show that emission from repopulated LE states accounts to not more than ~9% to the total emission. Moreover, any reformed LE states may dissociate again under the effect of a large internal field, which depletes the LE emission channel.

Figure **4a** shows the $J_{photo}$, $\Phi_{ph,max}$, and $Q_{extr}$ for PM6:Y5, all measured up to high $F_{eff}$. In addition, the $J_{photo}$ of the prototypical PM6:Y6 OPV is plotted for the same field range. Since both PM6:Y5 and PM6:Y6 blends exhibit similar absorption spectra, it must follow that the exciton generation rates in the two blends are comparable. This, paired with activation-less free charge generation exhibited by PM6:Y6, lets us presume that the reverse saturation of $J_{photo}$ in PM6:Y6 represents the situation where (nearly) all NFA excitons dissociate into CT states and finally into free charges, i.e. where free charge generation proceeds with nearly no activation barrier. We first note that at sufficiently high $F_{eff}$, the $J_{photo}$ of the poorly performing PM6:Y5 OPV approaches the $J_{SC}$ of PM6:Y6 ($J_{high-field,PM6:Y5}$=24.1 mA/cm$^2$ vs. $J_{SC,PM6:Y6}$=24.9 mA/cm$^2$). This means that almost all excitons generated in PM6:Y5 can eventually dissociate under a sufficiently high $F_{eff}$ and contribute to photocurrent. Note that the lack of saturation of $J_{photo}$ even at an $F_{eff}$ approaching $1 \times 10^8$ V.m$^{-1}$ suggests that a small barrier for LE or CT dissociation still exists under high fields. Furthermore, it turns out that $Q_{extr}$ from TDCF can be perfectly superimposed on $J_{photo}$ at high reverse bias (high $F_{eff}$); indicating the region where the field-dependence of $J_{photo}$ is completely described by the field-dependence of free charge generation, with vanishingly low free charge carrier recombination. We now correlate the $J_{photo}$ and $Q_{extr}$ data sets with

$\Phi_{ph,max}$ from field-dependent ssPL spectra. We find that (a) the $\Phi_{ph,max}$ of PM6:Y5 can be completely overlaid on the $Q_{extr}$ data over the entire field range and (b) that for this scaling a zero ssPL intensity corresponds to the reverse saturation photocurrent of PM6:Y6 – where there is no barrier for exciton dissociation. This explains the importance of the grey dashed horizontal line in Figure **4a**. The fact that there is little deviation between $\Phi_{ph,max}$ and $Q_{extr}$ even near $V_{OC}$, where photocurrent loss is mainly due to NGR, supports our early assignment that the ssPL has little contribution from exciton reformation due to free charge carrier recombination.

At this point, the question arises as to whether the observed field-dependence of LE dissociation is that of LE states generated in the bulk of neat Y5 domains or whether it is an interfacial phenomenon, i.e., it only helps to overcome a barrier towards CT formation. For this, we compare $\Phi_{ph,max}$ versus bias for PS:Y5 devices with that of PM6:Y5 devices with the same device geometry, shown in Figure **4a**. The peak PL photon flux is field-independent in the PS:Y5 case, which highlights the fact that that the effective internal field only affects exciton dissociation at the DA heterojunction, i.e., formation of interfacial CT states. Such a scenario requires that an appreciable energetic barrier exists between LE and CT states, which is indeed predicted by Marcus theory for the case of a small energetic offset and an appreciable reorganization energy.[23] This situation is depicted in Figure **4b**. The effect of $F_{eff}$ lowers the CT state energy which in turn reduces the barrier towards CT state formation. This would require the CT to carry a large electric dipole moment and/or to be highly polarizable. [24] Significant charge delocalization of the CT state has been predicted recently for the prototypical NFA Y6, which could explain a large dipole moment for the CT state. [25] Such effects could also be applicable to Y5-based blends, due to similar NFA chemical structure.

In conclusion, we studied the interrelation between free charge generation and exciton decay in optimized blends of the polymer donor PM6 with the Y5 NFA, a non-halogenated sibling of the NFA Y6 albeit exhibiting a lower HOMO-HOMO offset with the polymer donor. We developed a new TDCF method called m-TDCF to prove the absence of recombination-induced artefacts in the experimental free charge generation data. By systematically studying devices with different active layer thicknesses and with two different ETLs, the field-assisted process is revealed to be a bulk phenomenon. Finally, we show that the field dependencies of free charge generation and photoluminescence intensity are perfectly anticorrelated, with zero emission intensity corresponding to the case of complete exciton dissociation. These findings clearly point to insufficient LE dissociation as the limiting factor for photocurrent generation in this system. Our approach can be used to analyse the field-dependence of other NFA-based systems, which are currently being studied to obtain an eagle's eye view on free charge generation in low energetic offset NFA based OPVs.

## Acknowledgements

We thank the Deutsche Forschungsgemeinschaft (DFG, German Research Foundation) for funding through the project Extraordinaire (project number 460766640).

# Anticorrelated Photoluminescence and Free Charge Generation Proves Field-Assisted Exciton Dissociation in Low-Offset PM6:Y5 Organic Solar Cells

*Manasi Pranav, Thomas Hultzsch, Artem Musiienko, Bowen Sun, Atul Shukla, Frank Jaiser, Safa Shoaee, Dieter Neher*

Affiliations:

M. Pranav, T. Hultzsch, Dr. A. Shukla, Dr. F. Jaiser, Prof. D. Neher
Soft Matter Physics, Institute of Physics and Astronomy
University of Potsdam
Karl-Liebknecht-Str. 24-25, 14476 Potsdam-Golm, Germany
E-mail: neher@uni-potsdam.de

M. Pranav, B. Sun, S. Shoaee
Optoelectronics of Disordered Semiconductors, Institute of Physics and Astronomy
University of Potsdam
Karl-Liebknecht-Str. 24-25, 14476 Potsdam-Golm, Germany

Dr. A. Musiienko
Department Novel Materials and Interfaces for Photovoltaic Solar Cells,
Helmholtz-Zentrum Berlin für Materialien und Energie,
Keküléstraße 5, 12489 Berlin, Germany

## Experimental Methods:

### Device fabrication:

The polymer PM6 and the small non-fullerene acceptor Y5 were both purchased from 1-Materials Inc. The solvents chloroform ($CHCl_3$) and the methanol (MeOH) were purchased from Carl Roth and Sigma Aldrich, respectively. Optimized devices were fabricated in a conventional geometry with a structure ITO/PEDOT:PSS/PM6:Y5/PDINO or PDINN/Ag. Glass substrates with pre-patterned ITO (Lumtec) were cleaned in an ultrasonic bath with acetone, Hellmanex III, deionized water and isopropanol for 10 minutes each, followed by microwave oxygen plasma treatment (4 min at 200 W). Subsequently, an aqueous solution of PEDOT:PSS (Heraeus Clevios™ PEDOT:PSS) was filtered through a 0.2 µm PTFE filter and spin coated onto ITO at 5000 rpm under ambient conditions. The PEDOT:PSS coated substrates were thermally annealed at 150°C for 25 min. PM6 and Y6 were dissolved in CHCl3 to a total concentration of 14 $mgmL^{-1}$ with a 1:1.2 weight ratio with no additive. The solution was stirred for 3 hours at 40°C inside a nitrogen-filled glovebox. Polystyrene (Sigma Aldrich) blends with Y5 were prepared similarly, with a 14$mgmL^{-1}$ concentration and a 1:19 ratio. The blend was spin coated at 2000rpm onto the PEDOT:PSS coated substrates to obtain a photoactive layer of thickness of 100 nm. For 50 nm, 150 nm, and 250 nm active layer thicknesses, a PM6:Y5 blend solution of concentration 9 $mgmL^{-1}$, 14 $mgmL^{-1}$, 18 $mgmL^{-1}$, respectively was spin coated at 3500 rpm, 1200 rpm and 1000 rpm respectively. Then, a 1 $mgmL^{-1}$ solution of PDINO in methanol was spin coated at 2000 rpm (1500 rpm for a 1 $mgmL^{-1}$ solution of PDINN). Finally, 100 nm of Ag as the top electrode was evaporated under a

$10^{-6}$-$10^{-7}$ mbar vacuum. The resulting active area was 1 mm$^2$ for TDCF and PL measurements and 6 mm$^2$ for EQE and JV measurements.

### *Current density-voltage characteristics (JV)*

JV curves were measured using a Keithley 2400 SourceMeter in a 2-wire source configuration. Simulated AM1.5G irradiation at 100 mWcm-2 was provided by a filtered Oriel Sol2A Class AA Xenon lamp and the intensity was monitored simultaneously with a Si photodiode. The sun simulator is calibrated with a KG5 filtered silicon solar cell (certified by Fraunhofer ISE).

### *EQE$_{PV}$ and absorbance*

The EQEPV was measured with broad white light from a 300 W Halogen lamp (Phillips) which was chopped at 80 Hz (Thorlabs MC2000), guided through a Tornerstone monochromator and coupled into an optical quartz fibre, calibrated with Newport Photodiode (818-UV). An SR 830 Lock-In Amplifier measures the response of the solar cell under different bias voltages applied by a Keithley 2400.

Absorbance was measured on films coated under same fabrication conditions mentioned above, but on glass substrates with a Varian Cary 5000 spectrophotometer in transmission mode.

### *Bias dependent photoluminescence (PL) and absolute PL/EL (PLQY/ELQY)*

Bias-dependent PL measurements were performed using a 520 nm CW laser diode (Insaneware) for steady state illumination, and the intensity of the laser was adjusted to a 1 sun equivalent by illuminating a PM6:Y6 solar cell under short-circuit (provided by a Keithley 2400) and matching the current density reading to the $J_{SC}$ obtained in the sun simulator. The excitation beam was focused onto the sample using a stage of mirrors and lenses. Bias voltages ranging from open-circuit voltage to -8 V were applied to the sample using the same Keithley 2400. To ensure that only the active layer is illuminated and contributes to the emission response, we masked the measured pixels. The emission spectra were recorded with an Andor Solis SR393i-B spectrograph with a silicon DU420A-BR-DD detector and an Indium Gallium Arsenide DU491A-1.7 detector. A calibrated Oriel 63355 lamp was used to correct the spectral response. PL spectra were recorded with different gratings at central wavelengths of 800, 1100, and 1400 nm, and merged afterwards. For PLQY measurements, the same laser was used but the excitation beam was channelled through an optical fibre into an integrating sphere containing the sample. A second optical fibre was used from the output of the integrating sphere to the Andor Solis SR393i-B spectrograph. The spectral photon density was obtained from the corrected detector signal (spectral irradiance) by division through the photon energy ($h\nu$), and the photon numbers were calculated from the numerical integration, using a Matlab code.

For absolute EL measurements, a calibrated Si photodetector (Newport) connected to a Keithley 485 picoampere meter was used. The detector, with an active area of ~2 cm$^2$, was placed in front of the measured pixel with a distance <0.5 cm, and the total photon flux was evaluated considering the emission spectrum of the device and the external quantum efficiency of the detector. The injected current was monitored with a Keithley 2400.

*Time-delayed collection field (TDCF) and modified-TDCF (mTDCF)*

In TDCF, the device was excited with a laser pulse from a diode pumped, Q-switched Nd:YAG laser (NT242, EKSPLA) with ~6 ns pulse duration at a typical repetition rate of 500 Hz. To compensate for the internal latency of the pulse generator, the laser pulse was delayed and homogeneously scattered in an 85 m long silica fiber (LEONI) after triggering a photodiode. An Agilent 81150A pulse generator was used to apply a square voltage transient waveform constituting the pre-bias $V_{pre}$ and collection bias $V_{coll}$. The device was illuminated while held at different pre-bias $V_{pre}$. After a pre-set delay time (calculated from the falling slope of the photodiode trigger), a high reverse bias $V_{coll}$ lower than the minimum $V_{pre}$ was applied to extract all the charges generated in the device. $V_{pre}$ and $V_{coll}$ were set by the Agilent 81150A pulse generator through a home-built current amplifier, which was triggered by a fast photodiode (EOT, ET-2030TTL). The current flowing through the device was measured via a 10 Ω resistor in series with the sample and recorded with an oscilloscope (Agilent DSO9104H). To avoid non-geminate recombination of photogenerated free charge carriers prior to extraction, the intensity of light is kept very low and the delay time of collection is set to ~1ns.

For mTDCF, a square-type waveform was programmed and fed into the Agilent 81150A pulse generator, and the delay parameters for the $V_{pre}$ and $V_{coll}$ voltage steps were pre-defined prior to the measurement. As in regular TDCF, the laser is delayed to compensate for the internal latency of the function generator.

*AC Hall-effect measurements*

Hall-effect with alternating field (AC Hall) measurements were conducted with an 8400 series of tools manufactured by Lake Shore Cryotronics and operated with 100 MHz and 0.6 T amplitude magnetic field. The conductivity, resistivity, carrier concentration, and Hall mobility parameters were directly measured using 4-probe Van der Pauw contact geometry.

## Supplementary data

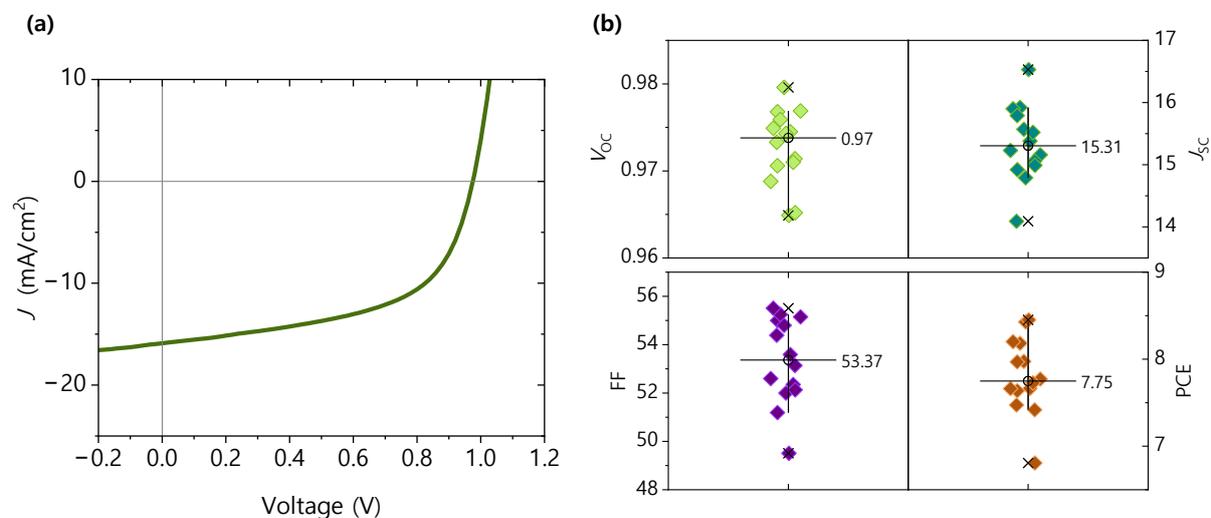

***Figure S1***. *(a) Current-voltage (JV) curve of an optimized PM6:Y5 OPV. (b) Statistics of the photovoltaic parameters for PM6:Y5 OPVs, obtained from JV measurements on 15 devices of 0.06cm$^2$ device area.*

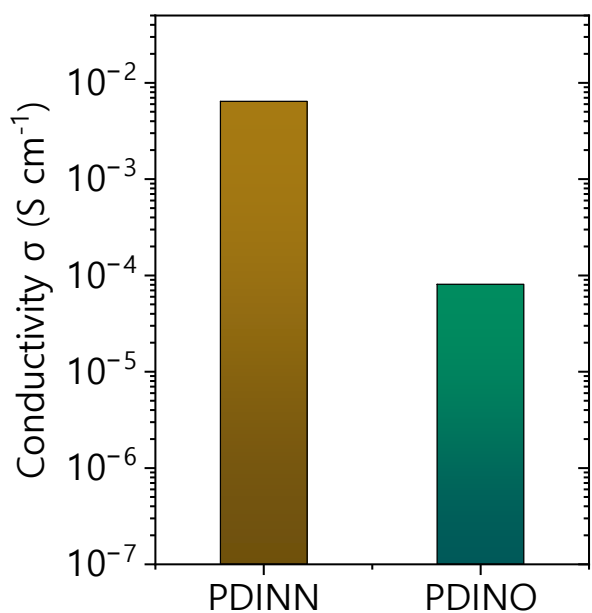

***Figure S2***. *Conductivity of two perylene diimide based ETLs with amino N-oxide terminal group (PDINO) and aliphatic amine terminal groups (PDINN). PDINN possesses two orders of magnitude higher conductivity than PDINO. Consequently, the voltage drop across the ETL is smaller and the dielectric relaxation time is shortened, which allows external biases across the OPV to be quickly applied to the active layer of the device.*

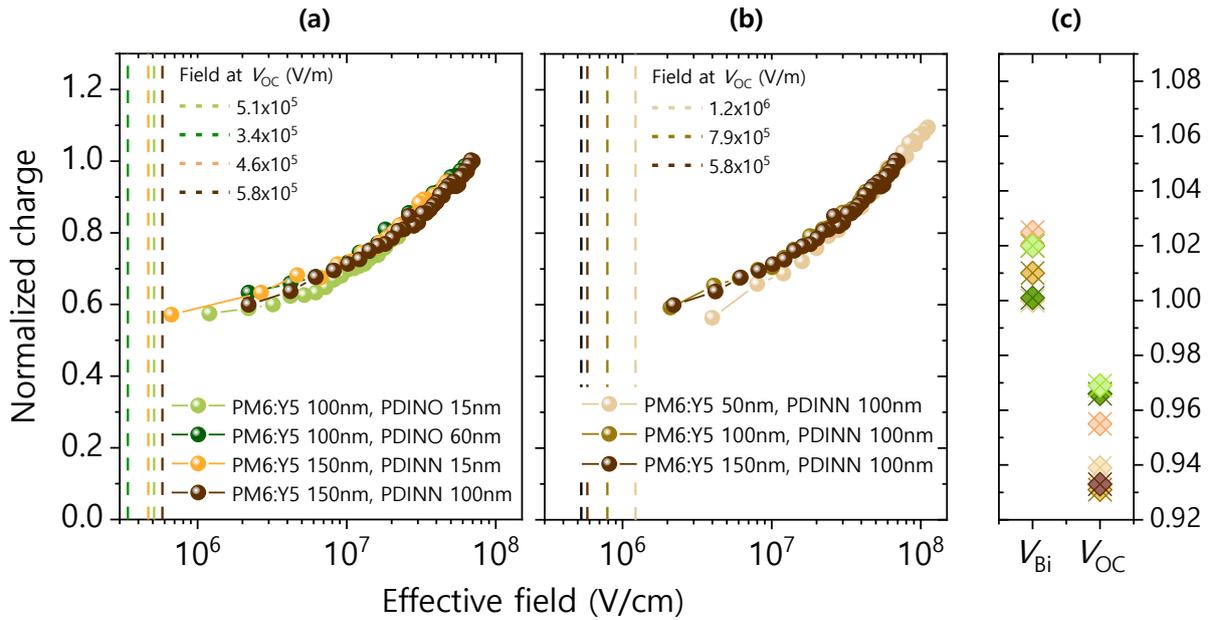

*Figure S3*. Semi-log representation of extracted charge $Q_{extr}$ that is normalised to an applied field of $7\times10^7$ V.m$^{-1}$ for (a) different ETL materials and thicknesses that were spin-coated over the PM6:Y5 layer and (b) for various PM6:Y5 layer thicknesses with a very thick PDINN ETL. The vertical dotted lines correspond to the effective field value at open-circuit. The curves perfectly overlap across the entire field range. We, however, note a slightly smaller collected charge for the devices with the thinnest active layer (50 nm) when approaching $V_{OC}$ in (b), which hints at non-geminate recombination of photogenerated and injected charges. (c) shows the built-in and open-circuit voltages for the measured devices, the former of which was used to translate the pre-bias voltage into an effective field.

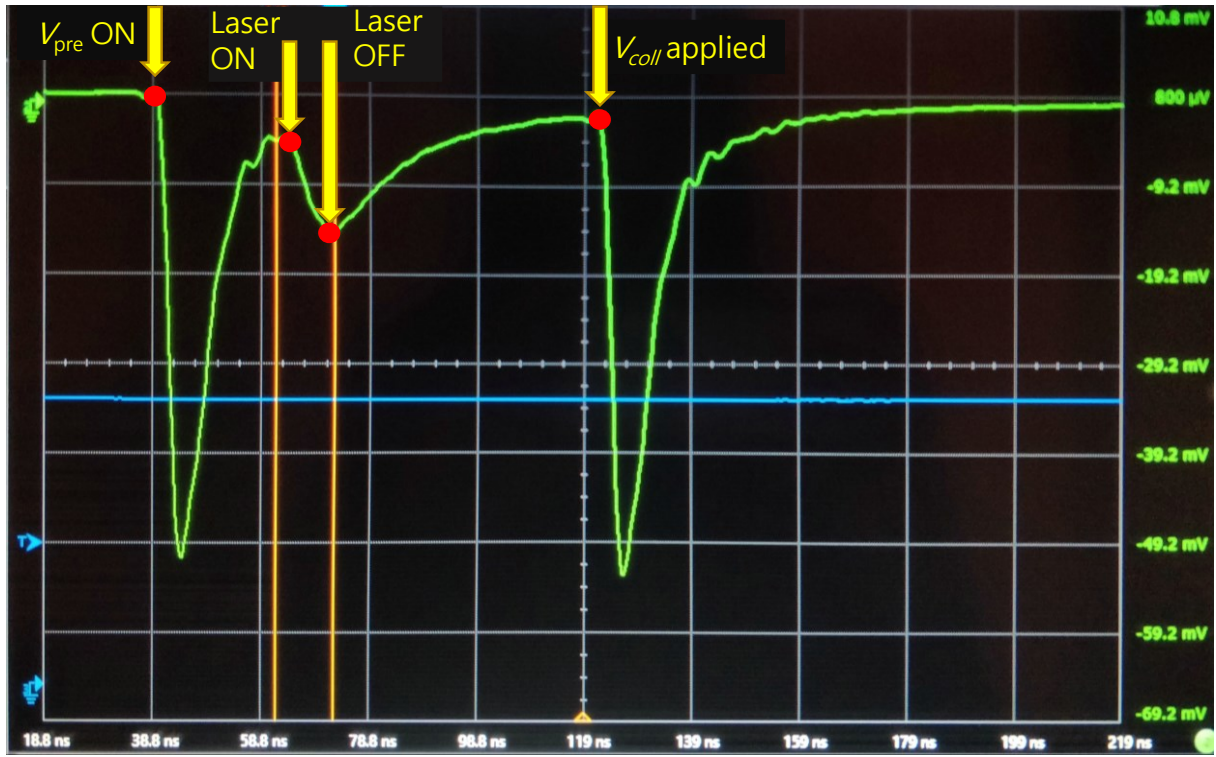

*Figure S4. Snapshot of the oscilloscope screen showing the photogenerated current trace (green line) in modified-TDCF, depicting the device's responses to the application of pre-bias $V_{pre}$, the laser and the extraction bias $V_{coll}$. Here, the scenario corresponds to a 50 ns delayed collection of free charge carriers, while the $V_{pre}$ is applied ~24ns before the laser pulse. The rise time of the photocurrent upon application of $V_{pre}$ and $V_{coll}$ is mainly limited by the slew rate of the function generation. This device had an RC time of 7 ns, meaning that ca. 21 ns after application of the pre-bias, the device would be charged to 95 % of the full external pre-bias voltage.*

In order to determine the optimum pre-bias duration $t_{adv,pre}$, the RC response time of the sample and the latency of the function generator need to be considered. Figure S4 shows a photocurrent transient for $V_{pre}$=-2V and $V_{pre}$=-4V. While the photocurrent rise is mainly limited by the rise time of the function generator, the initial current decay is determined by the RC time of the sample. To determine the RC time, we consider the dark equivalent of the photocurrent transient shown above and fit the decay of the pre-bias dark current transient with an exponential decay function. This yields an RC time of ca. 7ns. This agrees with a calculated RC time of ca. 7.7 ns for a device capacitance of 310 pF (assuming a dielectric constant of 3.5, device area of 1.1mm$^2$ and thickness of 110 nm) and a series resistance of ca. 30 Ω (arising from the ITO layer and input impedance of the current amplifier). After the initial decay, the dark current transient will assume a constant value, where the external voltage is fully applied across the capacitor and the current is determined by the continuous injection of charges into the active layer. Since the capacitor will take thrice the RC time to reach the point where it is charged up to 95% of the applied voltage, we take (3 x 7 ns) + ca. 3 ns (rise time of the function generator) = 24 ns as the delay between the onset of $V_{pre}$ and the onset of photoexcitation.

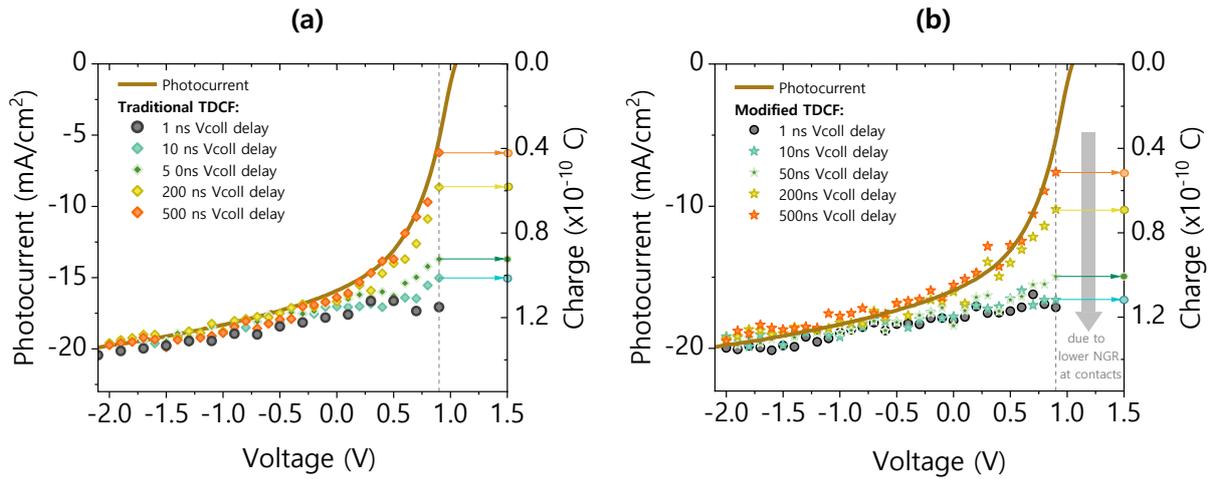

*Figure S5*. *An overlay of JV photocurrent of PM6:Y5 and bias-dependent extracted* charge *obtained with (a) traditional TDCF and (b) modified TDCF (m-TDCF), respectively, excited with a 2.33 eV laser pulse of 0.06 μJ/cm$^2$ fluence. In traditional TDCF, the pre-bias $V_{pre}$ is applied for 1.995 ms (2 ms minus the collection duration), whereas in m-TDCF, the pre-bias $V_{pre}$ is applied for 24 ns prior to photoexcitation. For longer extraction delays, the charge collected at negative biases is the same in TDCF and m-TDCF, but the charge collected near $V_{OC}$ (the vertical dotted line) is higher in m-TDCF, highlighted by the downward shift of the coloured circles (ordinates) on the two right-y-axes. This is evidence for the working principle of m-TDCF: reduced non-geminate recombination between photogenerated charge carriers and dark-injected charge carriers in the vicinity of the electrode.*

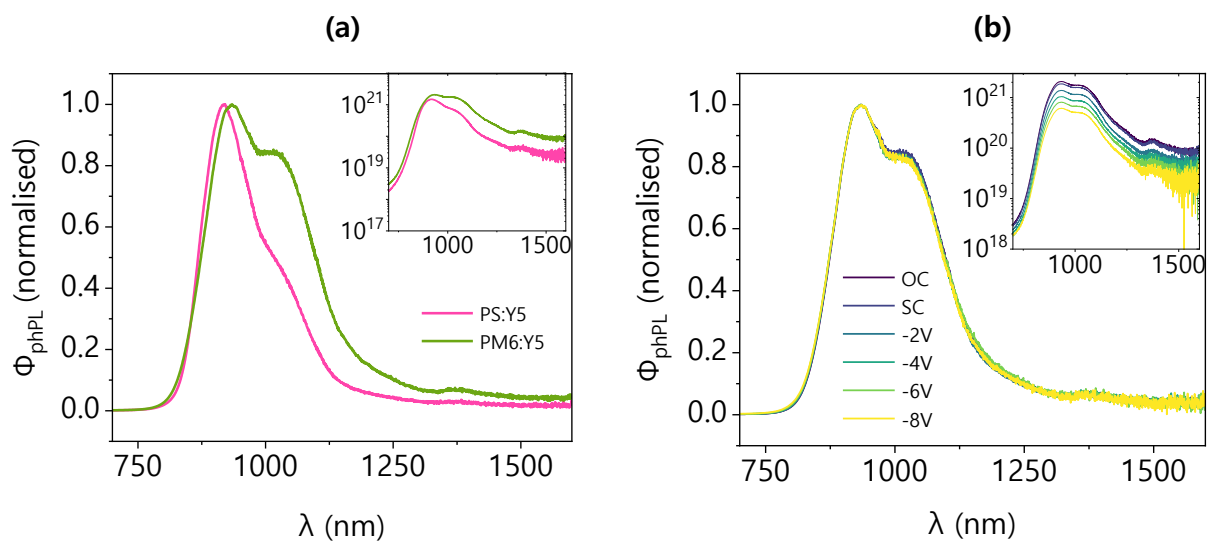

*Figure S6. (a) Normalised PL photon flux spectra for a polystyrene:Y5 blend device and a PM6:Y5 blend device measured at open-circuit. Though the PM6:Y5 emission is marginally red-shifted indicating slight changes of the Y5 aggregation, the spectral signature of the blend is largely governed by the emission from the neat acceptor. (inset) PL spectra for the two cases plotted on a semi-log scale. (b) Normalised spectra of emitted PL photon flux for ITO/PEDOT:PSS/PM6:Y5/PDINN/Ag as a function of external bias during measurement. The applied reverse bias does not change the spectral signature of PL emission, although the intensity is quenched due to enhanced exciton dissociation (inset in (b)).*